\documentclass[fleqn,usenatbib]{mnras}

\usepackage[T1]{fontenc}
\usepackage{ae,aecompl}

\usepackage{graphicx}	
\usepackage{amsmath}	
\usepackage{atbegshi}

\usepackage{booktabs} 
\usepackage{rotating}

\usepackage{txfonts}

\title[Rotating Globular Clusters from the SDSS-IV APOGEE-2 Survey]{The Rotation of Selected Globular Clusters and the Differential Rotation of M3 in Multiple Populations from the SDSS-IV APOGEE-2 Survey}

\author[Szigeti et al.]{
L{\'a}szl{\'o} Szigeti,$^{1}$\thanks{E-mail: szilac@gothard.hu}
Szabolcs~M{\'e}sz{\'a}ros,$^{1,2}$
Gyula M. Szab{\'o},$^{1}$
Jos{\'e} G. Fern{\'a}ndez-Trincado$^{3,4,5}$
\newauthor
Richard R. Lane,$^{3,5}$
Roger E. Cohen,${^6}$\\
$^{1}$ELTE Gothard Astrophysical Observatory, H-9704 Szombathely, Szent Imre Herceg st. 112, Hungary\\
$^{2}$Premium Postdoctoral Fellow of the Hungarian Academy of Sciences\\
${^3}$Instituto de Astronom{\'i}a y Ciencias Planetarias, Universidad de Atacama, Copayapu 485, Copiap{\'o}, Chile\\
$^{4}$Institut Utinam, CNRS UMR 6213, Universit\'e Bourgogne-Franche-Comt\'e, OSU THETA Franche-Comt\'e, Observatoire de Besan\c{c}on, \\    BP 1615, 25010 Besan\c{c}on Cedex, France \\
$^{5}$Centro de Investigaci\'on en Astronom\'ia, Universidad Bernardo O Higgins, Avenida Viel 1497, Santiago, Chile\\
${^6}$Space Telescope Science Institute, 3700 San Martin Drive, Baltimore, MD 21218, USA\\
}

\date{Accepted XXX. Received YYY; in original form ZZZ}

\pubyear{2021}

\begin{document}
\label{firstpage}
\pagerange{\pageref{firstpage}--\pageref{lastpage}}
\maketitle
\begin{abstract}
In this paper, we analyze 10 globular clusters in order to measure their rotational properties by using high precision radial velocity data from the SDSS-IV APOGEE-2 survey. Out of the 10 clusters we were able to successfully measure the rotation speed and position angle of the rotation axis for 9 clusters (M2, M3, M5, M12, M13, M15, M53, M92, M107). The comparison between our results and previous ones shows a really good agreement within our uncertainties.
For four of the globular clusters, M3, M13, M5 and M15, we separated the sample into two generation of stars using their [Al/Fe] abundances and examined the kinematic features of these generations separately from one another. In case of M3, we found significant difference between the rotational properties of first and second populations, confirming for the first time the predictions of several numerical simulations from the literature. The other three clusters (M5, M13, M15) also show smaller deviation between the two groups of stars, but those deviations are comparable to our errors.      
\end{abstract}




\section{Introduction}
The structure of globular clusters is shaped by complex interaction between external (interaction with the host galaxy) and internal (for example, relaxation, core collapse, mass segregation) forces. \cite{King1966} successfully described the star density of globular clusters with assuming no rotation.
\citet{Tiongco2017} have suggested that the internal relaxing processes dissipate all the angular momentum on a long term in every globular cluster. However, growing number of  studies \citet{Lane2010, Bellazzini2012, Bianchini2013, Fabricius2014, Lardo2015, Kimmig2015, Boberg2017, Lee2017, Cordero2017, Kamann2018, Ferraro2018, Lanzoni2018, Bianchini2018} present evidence that a significant amount of internal rotation could still be observed in many globular clusters, however the observed rotational strengths are only a fraction of the initial ones \citep{Kamann2018, Bianchini2018}. These studies mostly utilize recently recorded high quality radial velocity and proper motion data.

The presence of internal rotation could introduce some problems with the theory of globular clusters formation and evolution. Some studies indicate that the remaining rotation accelerates the evolution and shapes the morphology of the cluster \citep{Einsel1999, Bianchini2013}. Others suggest that the present day rotation could be a remnant of a strong rotation during the early history of globular clusters  \citep{Vesperini2014, Mapelli2017}. 

Orbital motions may be isotropic even during the formation of GCs, for example \citet{Lahen2020} have argued that the massive clusters are isotropic already during their first 100 Myr after formation.
Moreover, as \citet{Lane2010} and \citet{Tiongco2017} has shown, the interaction between the cluster and the galactic tidal field combined within the internal dynamics could produce complex kinematical features, e.g. radial variation in the orientation of the rotational axis, and anomalous velocity dispersions. The observational evidence of rotating globular clusters is based on high precision radial velocity data \citep{Bellazzini2012, Bianchini2013, Lardo2015, Kimmig2015, Boberg2017,Lee2017, Cordero2017, Ferraro2018, Lanzoni2018,Cordoni2020a}, integral field unit (IFU) spectrograph  \citep{Fabricius2014, Kamann2018}, and proper motion measurements \citep{Massari2013, Bellini2017, Bianchini2018, Mastrobuono2020}. 

The other aspect of the globular cluster rotation is the presence of multiple generations of stars and their rotational properties. 
In the last decade, the existence of two or more distinct generations of stars in most globular clusters became well studied \citet{Carretta2009a, Carretta2009b, Piotto2015, Milone2017, Milone2018, Meszaros2020}, however understanding the formation of multiple population is still an astrophysical challenge. Multiple populations manifest in light-element abundance variations, second generation stars (SG) are enhanced in N, Na, Ca and Al and depleted in C, O and Mg, while the first generation stars (FG) are the opposite. 
Most of the theories agree, that the second generation stars formed out of the first generation's ejecta mixed with the original intracluster medium \citep{Decressin2007, Bekki2010, Denissenkov2014, Bekki2017}, but the exact process of this pollution is currently not known, and many observed processes can not be explained with this theory yet. 
There are other alternative explanations for this phenomena and they are discussed in detail in \citet{Henault2015, Bastian2018}. 

Observational evidence have showed differences in spatial distribution between FG and SG stars. For dynamically younger clusters, the SG is more concentrated than the FG \citep{Sollima2007, Bellini2009, Lardo2011, Cordero2014, Boberg2016, Lee2017, Gerber2020}. On the other hand, the two generations are completely mixed in globular clusters with more advanced evolutionary stages \citep{Dalessandro2014, Nardiello2015, Cordero2015, Gerber2018, Gerber2020}. In \citet{Dalessandro2019}, the link between concentration differences and evolutionary stage was explored in detail based on observations and models.
Other observational studies revealed differences in the kinematics between the multiple generations \citep{Richer2013, Bellini2015, Bellini2018, Cordero2017, Milone2018, Libralato2019, Cordoni2020a, Cordoni2020b}, in other cases, the multiple generation of stars share similar kinematic properties \citep{Pancino2007, Cordoni2020b}
These literature sources used different indicators of cluster kinematics, including rotation, velocity dispersion and anisotropy to show the different kinematical properties of multiple populations.

Our main purpose is to investigate the rotational properties of the selected clusters using high precision and homogeneous radial velocity data. Our secondary goal is to identify potential differences in the cluster rotation properties between the multiple populations.
For this study, we use state-of-the-art data from the high-resolution spectroscopic survey Apache Point Observatory Galactic Evolution Experiment (APOGEE) \citep{Majewski2016}. The APOGEE started as one component of the $3^{\rm rd}$ Sloan Digital 
Sky Survey (SDSS-III; \cite{Eisenstein2011}) and continues as part of SDSS$-$IV \citep{Blanton2017} as APOGEE-2\footnote{http://www.sdss.org/surveys/apogee-2/}. The goal of APOGEE-2 is to obtain high-resolution (R = 22500), high signal-to-noise, H-band spectra ($\lambda$ = 1.51$-$1.70$\mu$m) of more than 600,000 late-type stars in the Milky Way by the end of 2020, and to determine chemical abundances of $\sim$26 elements in all observed stars. 
Most APOGEE targets are evolved red-giant branch (RGB) and asymptotic giant branch (AGB) stars from all major Galactic stellar populations.


\section{Data and reduction}
\subsection{Target Selection and Radial Velocities}
 
The data were gathered by the Sloan Foundation 2.5 m Richey-Chreritien altitude-azimuth telescope \citep{Gunn2006} at Apache Point Observatory. The spectra were obtained via the APOGEE spectrograph \citep{Wilson2019} with a resolution power of 22500. The stellar atmospheric parameters and chemical abundances are calculated from these spectra with APOGEE Stellar Parameters and Chemical Abundances Pipeline (ASPCAP) \citep{Garcia2016}.    
We use the list of stars compiled by \citet{Masseron2019}. The target selection is explained in detail in \citet{Meszaros2015}, \citet{Masseron2019} and \citet{Meszaros2020}. In short, stars were selected based on their radial velocity, distance from cluster center and metallicity. In radial velocity, all studies required stars to be within three times the velocity dispersion of the mean cluster velocity, which were taken from \citet{Baumgardt2019}.

We used the DR14 data release of APOGEE \citep{Holtzman2018} for our study. Radial velocities are derived by the reduction pipeline \citep{Nidever2015}, while details can be found in that paper, we provide a brief description of the algorithm. For almost all stars, observations are made in multiple visits to improve S/N and to allow observations of faint objects, which are then combined together to provide the final spectrum of the star. The radial velocity is measured in multiple steps. First, we do an initial measurement for each star from the actual individual spectrum by cross correlating each spectrum with the best match in a template library. In the second step, all of the visits are combined, and relative radial velocities of each visit are iteratively refined by cross-correlating each visit spectrum with the combined spectrum. The final absolute radial velocity is set by cross-correlating the combined spectrum with the best match in a template library. The peak of the cross-correlation function is fitted with a Gaussian in order to determine the accurate spectral shift. 
Binary stars can distort the rotational velocity profile if included in the sample. We removed all the binaries from our sample using database from \citet{PriceWhelan2020}. In \citet{PriceWhelan2020}, they identified nearly 20000 binary candidates with high confidence in the 16th data release of APOGEE, out of these we have found 65 stars in common with our targets, only $\approx~7\%$ of the total number of stars.
 
The uncertainties of radial velocities depend on multiple factors, mainly the characteristic of spectra, the resolution and the S/N ratio. For example, a star with lots of deep and thin lines in the spectra had a more precise RV than a star with wide and shallow lines. The typical uncertainty of the final radial velocity for stars in this study is 0.1 km/s.

\subsection{Method}

\begin{table*}                                                                                 
\begin{tabular}{lccccccccc}                                                                                           
\toprule                                                                                                                
GC name & N & [Fe/H] & V$_{\rm helio}$ & R$_{\rm h}$ & d$_{\rm avg}$ & PA & PA$_{\rm err}$ & A$_{\rm rot}$ & A$_{\rm rot err}$  \\          
&  &       &   [km/s] & [arcmin]  &   [arcmin]   &       &            & [km/s]        & [km/s]                        \\    

\hline                                                                                                                
M2 & 26 & -1.65 & -5.3 & 0.93 & 3.9 & 26 & 19 & 3.48 & 0.82            	\\                        

M3 & 145 & -1.5 & -148.6 & 1.12 & 6.3 & 164 & 13 & 1.19 & 0.28          	\\

M5 & 215 & -1.29 & 52.1 & 2.11 & 5.9 & 148 & 6 & 3.45 & 0.37           	\\

M12 & 65 & -1.37 & -42.1 & 2.16 & 5.8 & 56 & 93 & 0.24 & 0.19                \\

M13 & 135 & -1.53 & -246.6 & 1.49 & 5.1 & 26 & 9 & 2.38 & 0.39  	        \\                        

M15 & 138 & -2.37 & -107.5 & 1.06 & 4.8 & 120 & 11 & 2.38 & 0.44      	\\                        

M53 & 40 & -1.86 & -79.1 & 1.11 & 4.8 & 98 & 27 & 1.54 & 0.57           	\\                        

M71 & 28 & -0.78 & -22.9 & 1.65 & 2.7 & ... & ... & ... & ...               	\\                        
 
M92 & 72 & -2.31 & -121.6 & 1.09 & 4.2 & 154 & 14 & 2.06 & 0.58       	\\                       

M107 & 67 & -1.02 & -33.8 & 2.70 & 4.3 & 168 & 30 & 0.72 & 0.27         	\\

\bottomrule 
\end{tabular}  
\caption{The table contains the basic parameters of the targeted globular clusters and the results of this study. The second column represents the number of observed stars. The third and the fourth columns are the metallicity and the clusters heliocentric radial velocity \citet{Harris1996, Miocchi2013}. The fifth is the halflight radii \citet{Harris1996}. The sixth is the average distance from the cluster centers in our samples. The seventh and eighth columns represent the calculated position angle and its error, while the last two columns are the rotational velocity and its uncertainties. The position angles measured from North to East anti-clockwise direction.}
\end{table*}

In order to investigate the rotational velocity and the position angle of the rotational axes of the cluster, we follow the same method as \cite{Cote1995, Bellazzini2012, Bianchini2013, Lardo2015, Kimmig2015, Boberg2017,Lee2017, Lanzoni2018}. 
First, each cluster was split into two halves along the cluster center. The position angle of separation was the independent variable of the analysis, varied between PA = 0 and 180 (PA = 90 is toward East) in 2 degree step-size.We ran the simulations with multiple step-sizes (e.g. 2, 5, 10, 20 degree) all providing the same end results within our uncertainties. At the end we choose the 2 degrees for all of our clusters to appropriately sample the densest areas.
Next, the mean radial velocity of these sub-samples were calculated and the difference between the two sub-samples mean velocity were determined. If rotation is present in the system, the delta V$_{\rm mean}$ draws a sinusoidal variation as a function of the position angle. The amplitude of the function is twice the rotational velocity (because the amplitude is the difference of the two hemispheres) and the min-max position (ideally the difference is 180 degree) is the PA of the rotational axis. 
Thus, we caution the reader that rotational velocity values printed in Table 2 and 3, and all figures are twice as large as the real rotational velocity, in agreement to what has been used in the literature.
Our results are listed in Table~1.

\subsection{Separating multiple populations in GCs}

In this study, first (FG) and second generation (SG) stars are separated from each other based on their [Al/Fe] abundances, following the suggested cuts by \citet{Meszaros2020} from APOGEE data. Previously, \citet{Meszaros2015} used an extreme-deconvolution code for fitting a distribution with multiple Gaussian components to identify FG from SG stars in a mutli-dimensional abundance space, they showed that almost all of the SG stars have [Al/Fe]$>$0.3~dex. For this reason we use this simplified criteria to set multiple populations apart using abundances from \citet{Masseron2019}. For this analysis, we selected 4 clusters with the most observed stars, in which we have enough samples to properly fit the rotational curve.

\subsection{Error estimation}
We checked the robustness of the method with a simple jackknife test. We randomly dropped more and more stars from the sample and calculated the rotational curves. The results indicate, we can get a good signal if we have at least 20-30 stars to work with. Fewer stars than this is insufficient for a robust measurement. 

In order to calculate the final uncertainty, we randomly dropped 20 percent of the stars (dropping more than this may result in fewer than 20 stars for some clusters) and derived the position angle and the rotational velocity in the sub-samples. 
We repeated this process 100 times, then the standard deviation of the sinusoidal fit was calculated for the position angles and rotational velocities for each of these sub-samples. The final uncertainties seen in Figure~1 are the average of the differences between the original fit and the sub-samples.

We were able to define the errors for all the selected clusters, however for M12, the combination of the small number of observed stars, low rotational signal, and the error estimation method produced a high uncertainty of the PA.   
For M71 we could not get a clear sinusoidal signal from the data at hand.  
Table~1 contains our derived results. 

We tested the robustness and the stability of the derived rotational amplitudes with a bootstrap analysis. For both populations in all clusters, 100 bootstrap distributions have been realised with redistributing the measured velocities randomly (with re-sampling allowed) among the field stars. These samples suffered a complete loss of any information on the rotation, and contained a "null signal". The distributions were then evaluated following exactly the same method as in the case of the observations, and we observed the best-fit amplitude of the inferred rotating model. This amplitude was considered as the upper limit of the rotation amplitudes if the clusters would not rotate, and the measured amplitudes were just a product of numerical fluctuations in the data distribution.
The standard distribution of the amplitudes in the bootstrap samples were in the range of 0.5--0.8 ~km/s, proving that the detection of the rotation of all examined clusters is indeed significant.


\section{Discussion}

\subsection{Comparison with literature}
\begin{table*}\centering
\begin{tabular}{@{}lcc@{}c@{}cc@{}c@{}cc@{}c@{}cc@{}c@{}cc@{}c@{}}
\toprule
& \multicolumn{2}{c}{M2} & \phantom{abc}& \multicolumn{2}{c}{M3} & \phantom{abc} & \multicolumn{2}{c}{M5} & \phantom{abc} & \multicolumn{2}{c}{M12} & \phantom{abc} & \multicolumn{2}{c}{M13} \\
\cmidrule{2-3} \cmidrule{5-6} \cmidrule{8-9} \cmidrule{11-12} \cmidrule{14-15} 
& PA & A$_{\rm rot}$ && PA & A$_{\rm rot}$ && PA & A$_{\rm rot}$ && PA & A$_{\rm rot}$ && PA & A$_{\rm rot}$  \\ 
\hline
\cite{Lane2010}       & ... & ... && ... & ... && ...  & ... && 40 & 0.15$\scriptsize\pm0.8$ && ...& ...       \\ 
\cite{Bellazzini2012} & ... & ... && ... & ... && 157   & 2.6$\scriptsize\pm0.5$ && ... & ... && ...  & ...      \\
\cite{Fabricius2014}  & ... & ... && 192 $\scriptsize\pm11.8$& ... && 149$\scriptsize\pm5.6$ & ... && 89$\scriptsize\pm19.3$& ... && 17$\scriptsize\pm7.8$& ...       \\ 
\citet{Kimmig2015}    & 53 & 4.7$\scriptsize\pm1.0$ && ... & 0.6$\scriptsize\pm1.0$ && ...  & 2.1$\scriptsize\pm0.7$ && ... & 0.2$\scriptsize\pm0.5$ && ... & ...       \\ 
\citet{Lee2017}       & ... & ... && ... & ... && 128  & 3.36$\scriptsize\pm0.7$ && ... & ... && ... & ...       \\ 
\cite{Cordero2017}    & ... & ... && ... & ... && ...  & ...  && ... & ... && 14$\scriptsize\pm19$  & 2.7$\scriptsize\pm0.9$   	    \\
\cite{Kamann2018}     & 41.7$\scriptsize\pm2.7$ &... && ... & ... &&144$\scriptsize\pm20.3$ & ...  && ... & ... && ... & ...  	    \\ 
\cite{Ferraro2018}    & ... & ... && 151 & 1.0 && ... & ...   && ... & ... && ... & ...   	    \\  
\cite{Lanzoni2018}    & ... & ... && ... & ... && 145 & 4.0   && ... & ... && ... & ...  	    \\
\cite{Sollima2019}    & 14$\scriptsize\pm12.1$ & 3.01$\scriptsize\pm0.7$ && ... & 1.75$\scriptsize\pm0.4$ && 132$\scriptsize\pm6$ & 4.11$\scriptsize\pm0.4$   && ... & 0.93$\scriptsize\pm0.4$ && 15$\scriptsize\pm14.2$ & 1.53$\scriptsize\pm0.6$  	    \\ 
\hline
this work             & $26\scriptsize\pm19$ & $3.48\scriptsize\pm0.8$ && $164\scriptsize\pm15$ & $1.19\scriptsize\pm0.3$ && $148\scriptsize\pm6  $ & $3.45\scriptsize\pm0.4$   && $56\scriptsize\pm93$ & $0.24\scriptsize\pm0.2$ &&  $26\scriptsize\pm9$ & $2.38\scriptsize\pm0.4$ \\  
\bottomrule
\end{tabular}
\caption{Comparison with literature. Part~1. \\
Position angles and rotational amplitudes from earlier studies. Since different conventions were followed, we convert the published result to PA 90 = East, anti-clockwise system. Literature source which did not use the double rotational velocity were converted to our system. The first sub-column represent the position angle of the rotational axis and the second is the rotational velocity in [km/s] } 
\end{table*}

\begin{table*}\centering
\begin{tabular}{@{}lcc@{}c@{}cc@{}ccc@{}c@{}cc@{}ccc@{}}
\toprule
& \multicolumn{2}{c}{M15} & \phantom{abc}& \multicolumn{2}{c}{M53} & \phantom{abc} & \multicolumn{2}{c}{M71} & \phantom{abc} & \multicolumn{2}{c}{M92} & \phantom{abc} & \multicolumn{2}{c}{M107} \\
\cmidrule{2-3} \cmidrule{5-6} \cmidrule{8-9} \cmidrule{11-12} \cmidrule{14-15} 
 & PA & A$_{\rm rot}$ && PA & A$_{\rm rot}$ && PA & A$_{\rm rot}$ && PA & A$_{\rm rot}$ && PA & A$_{\rm rot}$  \\ 
\hline
\cite{Lane2009}       & ... & ... && nf & nf && ...  & ... && ... & ... && ...& ...       \\ 
\cite{Bellazzini2012} & 110 & 3.8$\scriptsize\pm0.5$ && ... & ... && 163   & 1.3$\scriptsize\pm0.5$ && ... & ... && 84  & 2.9$\scriptsize\pm1.0$ \\
\citet{Bianchini2013} & 106$\scriptsize\pm1$ & 2.84&& ... & ... && ...   & ... && ... & ... && ... & ...  \\
\cite{Fabricius2014}  & ... & ... &&113$\scriptsize\pm19.2$& ... && ...   & ... && 99$\scriptsize\pm12.0$& ... && ... & ...  \\ 
\citet{Lardo2015}     & 120 & 3.63$\scriptsize\pm0.1$ && ... & ... && ...   & ... && ... & ... && ... & ...  \\
\citet{Kimmig2015}    & ... & 2.5$\scriptsize\pm0.8$ && ... & 0.4$\scriptsize\pm0.7$ && ...   & 0.4$\scriptsize\pm0.8$ && ... & 1.8$\scriptsize\pm0.8$ && ... & ...  \\ 
\citet{Boberg2017}    & ... & ... && 74 & 2.8 && ...   & ... && ... & ... && ... & ...  \\    
\cite{Kamann2018}     &151$\scriptsize\pm$10.4& ... && ... & ... && ...   & ... && ... & ... && ... & ...  \\ 
\cite{Ferraro2018}    & ... & ... && ... & ... && ...   & ... && ... & ... && 167 & 1.2  \\ 
\cite{Sollima2019}    & 128$\scriptsize\pm28.8$ & 3.29$\scriptsize\pm0.5$ && ... & ... && ... & ...   && ... & 1.46$\scriptsize\pm0.6$ && ... & ...  	    \\ 

\hline
this work             & $120\scriptsize\pm11$ & $2.38\scriptsize\pm0.4$ && $98\scriptsize\pm27$ & $1.54\scriptsize\pm0.6$ && ...   &  ...&& $154\scriptsize\pm14$ & $2.06\scriptsize\pm0.6$ && $168\scriptsize\pm30$ & $0.72\scriptsize\pm0.3$ 
\\  
\bottomrule
\end{tabular}
\caption{Comparison with literature. nf = not found evidence of rotation. Part~2}
\end{table*}

The latest results available in the literature were collected in Table 2 and 3, which contain the calculated position angles of the rotational axis for clusters in common with our sample. These data also represented in Figure 3. There are multiple conventions used in the literature for angle and direction notations. We converted all these various approaches to the PA 90 = East convention. The last row contains our values.

We detected systematic rotation in almost all of the targeted globular clusters. We confidently could derive rotational velocity and position angles for nine out of the ten selected clusters. All nine clusters have been studied in the literature before, thus we are able to not only compare our results with previous studies, but also homogenize the rotational velocity as our radial velocities are from one homogeneous survey. 
We caution the reader that the assumption of a constant rotation velocity as a function of distance to the cluster centre is a significant simplification. 
The observations of \citet{Boberg2017, Bianchini2018, Sollima2019} have shown that the peak of the rotational curve is located approximately at the cluster half-light radius, however this location is expected to change during the evolution of the cluster \citep{Tiongco2017}.
We listed the halflight-radius and the average distance of our sample from the cluster centre in Table~1. In all cases, the average distance is at least 2-3 times larger than the halflight-radius suggesting that our assumption of a constant rotation velocity underestimates the magnitude of the rotational velocity.

Before such a comparison can be made one must transform the results from the literature to a common coordinate system (PA 90 = East, anti-clockwise). After the transformation we are able to conclude while for some of the cluster we have a good agreement within our uncertainties, other less studied clusters with fewer stars show larger than expected discrepancy between studies. In the next few sections we examine these differences for each cluster.   

\subsubsection{\rm M5}
M5 is a well observed cluster targeted by \cite{Bellazzini2012, Fabricius2014, Kimmig2015, Lee2017, Kamann2018,Lanzoni2018}, therefore it is an excellent object to use it as a standard to compare our results to, especially because these literature sources used different measurement methods.
The position angle varies between 144.3 to 157 degree in the literature, our result of 148 degree fits nicely in this picture. 

As mentioned before, our calculation technique is similar to many that studied M5 \citep{Bellazzini2012, Lee2017, Lanzoni2018}, and by comparing our results to these studies, we can find a good agreement in all cases. \citet{Fabricius2014} and \citet{Kamann2018} used IFU spectrograph for the analysis. The advantage of this method that it is possible to measure crowded stellar fields in order to perform a detailed analysis of dispersion fields and central rotation.   
In \cite{Bianchini2018} the rotational pattern was derived from proper motion data (GAIA). Despite the different observations and methodology we feel confident in our approach as it nicely reproduces the rotation velocity and angle reported by the mentioned studies for M5.
In \citet{Sollima2019} the rotation of M5 (among other clusters) was investigated via radial velocity component from VLT and Keck instruments and proper motion data from GAIA 2nd data release. In \citet{Sollima2019} the rotational velocity derived from all the three velocity components (taken into consideration the inclination of the rotational axis) while we were able to use only the line-of-sight part. Their derived results show a good agreement with ours in case of PA and the difference in rotational velocity can be explained by the fact that we observed line of sight velocity, while they were able to determine the inclination.  

\subsubsection{\rm M2}
\cite{Kimmig2015} derived the rotational velocity as A$_{\rm rot} = 4.5 ~\rm km/s$, which is slightly larger than our value at $3.48\pm0.8$. Considering our uncertainty we conclude that these differences are not substantial. The position angle also  differs from \citet{Kimmig2015, Kamann2018, Sollima2019}, but our limited sample size cause a 19 degree of uncertainty, which can explain the difference.

\subsubsection{\rm M3, M12 and M13}
We have a good amount of observed stars in M3 and the derived position angles are within errors to \cite{Ferraro2018}. The difference from \cite{Fabricius2014} can be explained by their relatively large errors and the different analysis method applied. 

Our derived result for the rotational properties of M12 have high uncertainty, probably because either the amplitude of the rotation is too small, or the inclination is close to 90 degrees, so that we look close to the direction of rotational axis. We have a good agreement with the results \cite{Lane2010} and considering the large uncertainty we agree well with \cite{Fabricius2014} too.   

Our results show a good agreement at the level of our uncertainties with \cite{Fabricius2014}, \cite{Cordero2017} and \citet{Sollima2019} for M13. The rotational amplitude is also really similar to the one presented in \cite{Cordero2017}. 

\subsubsection{\rm M15}
In case of M15 we have a slight deviation in the PA from one of the previous results, namely \cite{Kamann2018}, however we are in a good agreement with the other four \citep{Bellazzini2012, Bianchini2013, Lardo2015, Sollima2019}. The source of the misalignment with \cite{Kamann2018}, which is 30 degrees, might be due to two different reasons. One is that \cite{Kamann2018} used data from an IFU spectrograph and used Voronoi-binned maps of the mean velocity and velocity dispersion across the observed region.  The second reasons is that they focused on the cluster center region, while our sample contains fewer stars from the cluster's center and more from the outer region. There is a variation among the derived rotational amplitudes among all the literature sources, but there are on the level of what is expected from the uncertainties of the measurements and methods.  

\subsubsection{\rm M53}
For M53, our PAs value lies between \cite{Fabricius2014} and \cite{Boberg2017} with a relatively high error of margin. The cause of this probably is the low sample size and the relatively small rotational signal originated from the cluster. The problem caused by the usage of fewer observed stars in the calculation is visible in Figure~1. If the cluster contains less stars for this method, the signal become more scattered, however the rotation is still detectable in this case. In a study of \cite{Lane2009} a detectable rotational signal was not found for this cluster within the margin of error.  

\subsubsection{\rm M71}
In case of M71, we did not find conclusive fit for the available data, which is interesting considering that we selected 28 stars from M71, similarly to M2, but such a small sample size was enough for a measurement in that cluster. The lack of measurement may suggest the possibility of this globular cluster not rotating or perhaps the inclination of the axis is close to 90 degree or the rotational signal simply too small to be detectable from our sample. 

\subsubsection{\rm M92}
The difference between our PA and that of \citet{Fabricius2014} in M92 is significant. Since they have used a different calculation method, we might expect a large difference such as this, however in case of M5, M3 and M53 we have a good agreements with this source. For this reason we do not know why our PA differs so much from \citet{Fabricius2014}, especially that our sample is large enough for a reliable measurement. Our derived rotational amplitude is similar the one derived in \cite{Kimmig2015} and \cite{Sollima2019}. 

\subsubsection{\rm M107}
Our derived position angle is very close to \cite{Ferraro2018}, however there is a 90 degree of deviation from \cite{Bellazzini2012} and a significant difference between the rotational amplitudes too. We are not sure what causes this, because \cite{Bellazzini2012} used the same method as we did and for other clusters we found an acceptable agreement.

\begin{figure*}
\centering
\includegraphics[width=6.5in,angle=0]{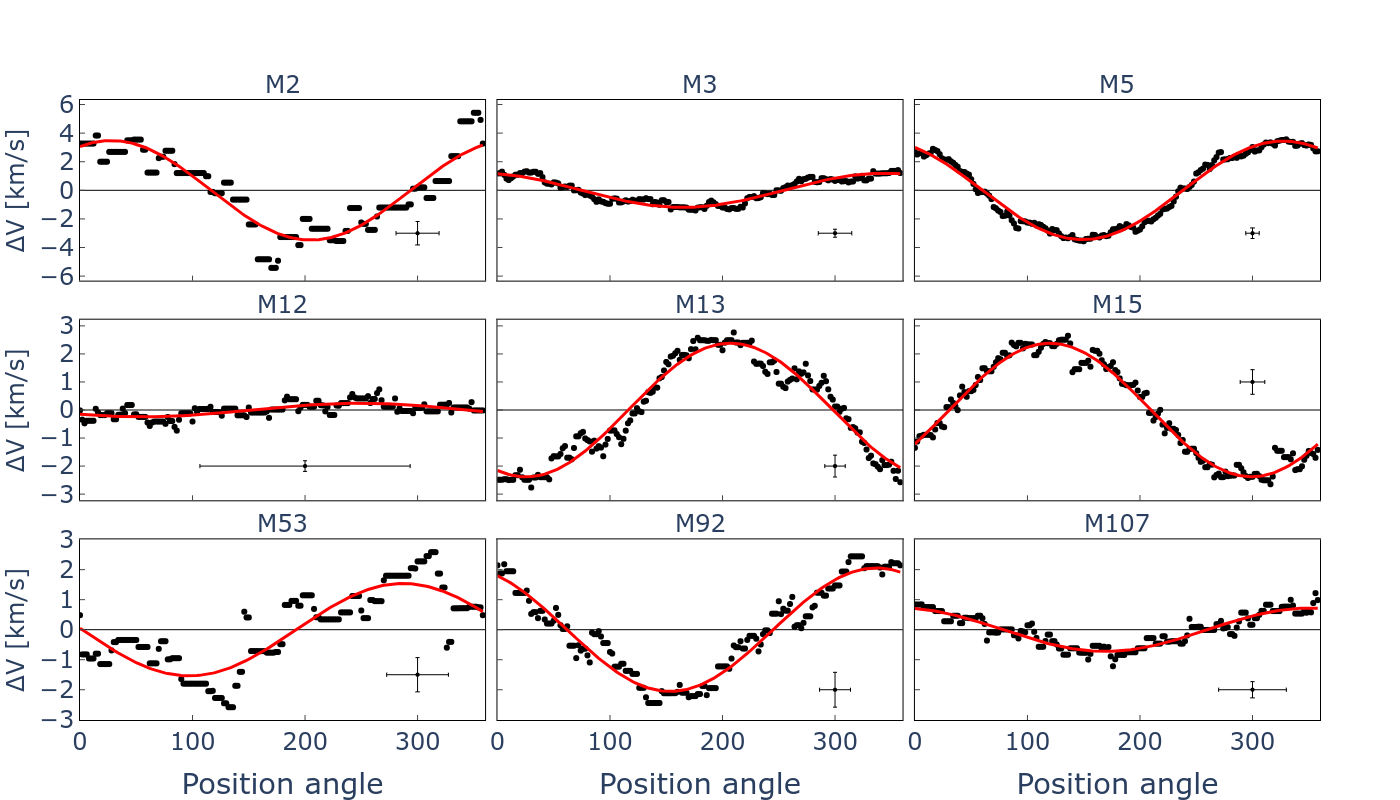}
\caption{ Global rotation of the globular clusters, having FG and SG samples unified. \\ 
Position angle of the rotational axis vs. difference between the two sub-sample's mean value in case all of the studied globular clusters except M71 (we can not find conclusive fit). The line shows the best fit sin function and the actual rotational velocity is half of the amplitudes. } 
\label{fig:m5radseb} 
\end{figure*}

\begin{figure*}
\centering
\includegraphics[width=6.5in,angle=0]{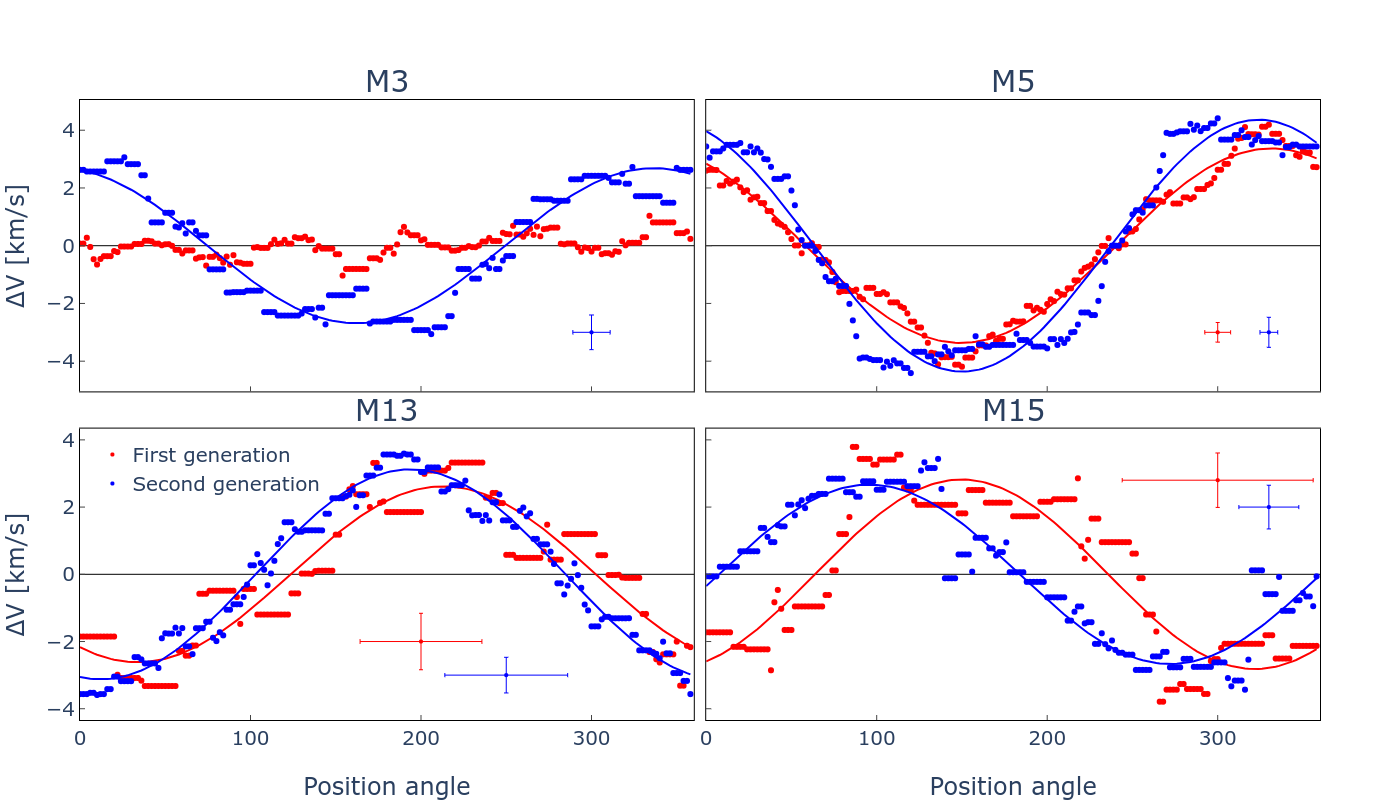}
\caption{Comparison of rotation curves of FG (red) and SG (blue) stars in the 4 selected globular clusters.    
} 
\label{fig:2gen_rvcurve} 
\end{figure*}

\begin{figure*}
\centering
\includegraphics[width=6.5in,angle=0]{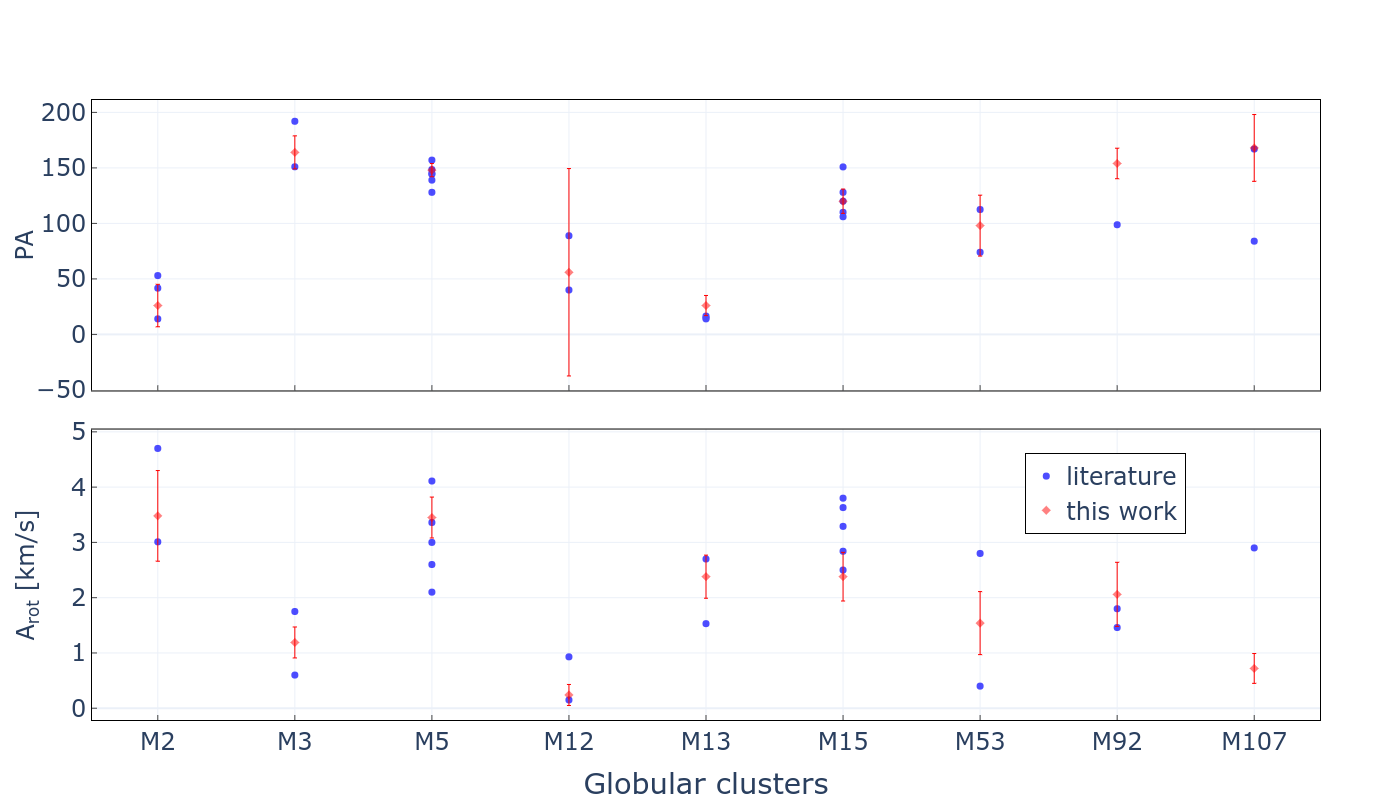}
\caption{Comparison between values determined in other studies (Table 2 and 3) and our results. 
} 
\label{fig:m5radseb} 
\end{figure*}

\subsection{Rotation according to first and second generation stars}

From theoretical studies \citep{Decressin2007, Bekki2010, Denissenkov2014, Bekki2017} we can expect GCs to have higher cluster rotational speed when measured from the SG stars than from only the FG stars. At the same time, the velocity dispersion should be higher among the FG stars. This is because the FG stars formed from massive molecular clouds and their first supernovas expelled the remaining cold gas from the GC. In a next stage, the polluted gas from the FG stars accumulated in the GCs center and this was the origin of the SG stars. Numerical simulations based on this theory \citep{Bekki2017} suggests higher rotational speed in case of SG than FG stars. We are able to test this idea in M3, M5, M13 and M15, clusters in which we have enough stars to sample both populations and are able to measure the cluster rotation based on the two generation of stars. Our results are listed in Table~4, and shown in Figure~2.  
However, we have to mention that in case of other formation scenarios opposite rotational velocities might be observed for FG and SG stars, i.e. the first generation rotates faster and the second slower as suggested by \citet{Henault2015}.    
Many different literature studies have examined the rotation as a function of stellar populations, such as 47 Tuc \citep{Milone2018}, Omega Centauri \citep{Bellini2018}, NGC 6352 \citep{Libralato2019}, M80 \citep{Kamann2020} or M54 \citep{Alfaro2020}, but none of these clusters are in our sample, thus no direct comparison is possible.

From figure~2 we can conclude that in case of M5, M13, M15 the rotational velocity originated from the FG and SG stars do not differ significantly, the small discrepancies are all well within our derived uncertainties. We are not able to measure the predicted difference in these three clusters. The position angles of M5 and M13 are also very close for both generation of stars, but we observe a large deviation in case of M15. The position angle is $144\pm56$ from the FG stars, and $98\pm18$ from the SG stars, but the extremely large uncertainty of the first value does not allow us to conclude that this difference have an astrophysical origin. A larger sample of observed stars and supplemented by proper motion data may shed some light in this phenomena. 

\citet{Cordoni2020a} examined the rotation of M5 according to FG and SG stars using GAIA proper motion data and line-of-sight velocities. The FG and SG stars present a significant difference in position angle in their study. We did not find these characteristics in our analysis, but it must be noted that our sample is much lower than theirs.

If we compare our M13 result (see in Table~4) to \cite{Cordero2017} in which the 'extreme' population correspond to SG with PA = 7 and the remaining stars ("normal" + "intermediate") to FG with PA = 33, then we have a really good agreement with our results( PA = 12 for SG and PA = 34 for FG). Although the uncertainties in both studies can be considered high, the observed differences are well within these errors. Considering our errors, we do not believe the discrepancy between the PA of the FG and SG group has an astrophysical origin.   

M3 is the peculiar object in our sample. Our observations prove that M3 does not appear to show any detectable global rotation when examined through only the FG group of stars, however in case of SG sample, we clearly see a strong rotation curves with an amplitude of $2.69\pm0.6$ km/s. The difference in cluster rotational velocities in the two population of stars is significant when compared to the uncertainties. 

In order to estimate the upper limit for a possible rotation that could remain hidden in the numerical fluctuations we used a bootstrap analysis,  described in Section 2.4. In M3, the standard deviation of bootstrap amplitudes was 0.55~km/s for FG stars, therefore the rotational amplitude is $<1.65$~km/s with a 3-$\sigma$ confidence. 

Our result appear to follow the theoretical predictions by \cite{Bekki2017}; the cluster rotational velocity based on the SG stars is significantly higher than based on the FG stars. The behavior of the FG stars is also interesting, because our observations suggest a rotational velocity very close to zero without any hint to what the PA might be. The simplest explanation is that the rotational velocity is so small that it is not possible to detect within our precision. 

\begin{table*}                                                                                 
\begin{tabular}{lcccccc}                                                                                    
\toprule
GC  & Gen &  N & PA  & A$_{\rm rot}$   \\          
\hline
M3 & (fg) & 95 & ... & ...                       \\
   & (sg) & 45 & 162 $\pm$ 11 & 2.69 $\pm$ 0.6         \\

M5 & (fg) & 102 & 150 $\pm$ 8 & 3.37 $\pm$ 0.3  	    \\
   & (sg) & 92 & 150 $\pm$ 6  & 4.36 $\pm$ 0.5        \\
   
M13 & (fg) & 36 & 34 $\pm$ 36  & 2.62 $\pm$ 0.8        \\                        
    & (sg) & 70 & 12 $\pm$ 36  & 3.12 $\pm$ 0.5       \\

M15 & (fg) & 33 & 144 $\pm$ 56  & 2.82 $\pm$ 0.8     	\\                        
    & (sg) & 49 & 98 $\pm$ 18   & 2.67 $\pm$ 0.7         \\
\bottomrule
\end{tabular}  
\caption{The first and second generation's kinematic properties in case of the four selected clusters.}
\end{table*}
  

\section{Conclusions}

We found evidence of rotation in M2, M3, M5, M12, M13, M15, M53, M92, and M107, but not in M71, supporting the theory that most globular cluster preserve significant amount of rotation during their lifetime. 
For most clusters, these results show good agreement with other similar studies. With the precise radial velocity data of the APOGEE survey, we were able to provide homogeneous rotational velocities and PAs for several clusters for which such homogeneity did not exist in independent literature sources. 

We successfully identified rotational signals originated from two different generation of stars in 4 selected clusters. 
In M3, we discovered a significant difference between the rotational velocity of the cluster when it is examined through only the FG and SG stars. This is very much in agreement with the prediction of numerical simulations by several independent groups \citep{Decressin2007, Bekki2010, Denissenkov2014, Bekki2017}. The cluster do not show a detectable rotational signal when selection the FG stars only,  while in case of the SG stars the cluster have a clear rotational signal at 2.69 km/s. It is not clear what cause this phenomena, but a detailed analysis with supplement proper motion data might unfold this issue. 

In M5 and M13 the FG rotational velocity is somewhat smaller than the SG velocity as the theory predicts, however the differences are well within the level or our uncertainties, thus we do not believe we see the prediction of the numerical simulations. In terms of PA, the deviations between the populations are really small or nonexistent and well within the derived uncertainties. 
For M15 one can see a difference in PAs, but the relatively high uncertainty of the PA for the FG stars prevents us to draw a clear conclusion, therefore further analysis with a larger sample size is required.

\section*{Acknowledgements} 

L. Sz. and Sz. M. has been supported by the Hungarian 
NKFI Grants K-119517 and GINOP-2.3.2-15-2016-00003 of the Hungarian National Research, Development and Innovation Office. Sz. M. has been supported by the J{\'a}nos Bolyai Research Scholarship of the Hungarian Academy of Sciences, and by the {\'U}NKP-20-4 New National Excellence Program of the Ministry for Innovation and Technology.

Funding for the Sloan Digital Sky Survey IV has been provided by the Alfred P. Sloan Foundation, 
the U.S. Department of Energy Office of Science, and the Participating Institutions. SDSS-IV acknowledges
support and resources from the Center for High-Performance Computing at
the University of Utah. The SDSS web site is www.sdss.org.

SDSS-IV is managed by the Astrophysical Research Consortium for the 
Participating Institutions of the SDSS Collaboration including the 
Brazilian Participation Group, the Carnegie Institution for Science, 
Carnegie Mellon University, the Chilean Participation Group, the French Participation Group, 
Harvard-Smithsonian Center for Astrophysics, 
Instituto de Astrof\'isica de Canarias, The Johns Hopkins University, 
Kavli Institute for the Physics and Mathematics of the Universe (IPMU) / 
University of Tokyo, Lawrence Berkeley National Laboratory, 
Leibniz Institut f\"ur Astrophysik Potsdam (AIP),  
Max-Planck-Institut f\"ur Astronomie (MPIA Heidelberg), 
Max-Planck-Institut f\"ur Astrophysik (MPA Garching), 
Max-Planck-Institut f\"ur Extraterrestrische Physik (MPE), 
National Astronomical Observatories of China, New Mexico State University, 
New York University, University of Notre Dame, 
Observat\'ario Nacional / MCTI, The Ohio State University, 
Pennsylvania State University, Shanghai Astronomical Observatory, 
United Kingdom Participation Group,
Universidad Nacional Aut\'onoma de M\'exico, University of Arizona, 
University of Colorado Boulder, University of Oxford, University of Portsmouth, 
University of Utah, University of Virginia, University of Washington, University of Wisconsin, 
Vanderbilt University, and Yale University.

\section*{Data availability} 
The data used in this article is part of the 14th Data Release from SDSS-IV / APOGEE survey, and it is publicly available at \url{https://www.sdss.org/dr14/data_access/}.



\bsp	
\label{lastpage}
\end{document}